\documentclass[preprint,12pt]{elsarticle}

\usepackage{graphicx}
\usepackage{comment}
\usepackage{float}
\usepackage{physics}
\usepackage{color}
\usepackage{overpic}
\usepackage{amssymb}
\newcommand{\beq}{\begin{equation}}
\newcommand{\eeq}{\end{equation}}
\newcommand{\lb}{\left(}
\newcommand{\rb}{\right)}

\usepackage{geometry}
\usepackage{amsrefs}
\usepackage{todonotes}
\journal{Journal Name}

\begin{document}

\begin{frontmatter}

%% Title, authors and addresses

\title{Predicting lift-off time when deep-frying potato dough snacks}

%% use the tnoteref command within \title for footnotes;
%% use the tnotetext command for the associated footnote;
%% use the fnref command within \author or \address for footnotes;
%% use the fntext command for the associated footnote;
%% use the corref command within \author for corresponding author footnotes;
%% use the cortext command for the associated footnote;
%% use the ead command for the email address,
%% and the form \ead[url] for the home page:
%%
%% \title{Title\tnoteref{label1}}
%% \tnotetext[label1]{}
%% \author{Name\corref{cor1}\fnref{label2}}
%% \ead{email address}
%% \ead[url]{home page}
%% \fntext[label2]{}
%% \cortext[cor1]{}
%% \address{Address\fnref{label3}}
%% \fntext[label3]{}

%% use optional labels to link authors explicitly to addresses:
%% \author[label1,label2]{<author name>}
%% \address[label1]{<address>}
%% \address[label2]{<address>}

\author{\footnote{All authors contributed equally to this work, and are placed in alphabetical order.}T Babb$^a$, GP Benham$^b$, R Gonzalez-Farina$^a$, KB Kiradjiev$^a$, WT Lee$^c$, S Tibos$^d$}

\address{a. Mathematical Institute, University of Oxford, Andrew Wiles Building, Radcliffe Observatory Quarter, Woodstock Road, Oxford OX2 6GG, United Kingdom\\
b. LadHyX, UMR CNRS 7646, Ecole polytechnique, 91128 Palaiseau, France\\
c.  School of Computing and Engineering, University of Huddersfield, Queensgate, Huddersfield HD1 3DH, United Kingdom. MACSI, Department of Mathematics and Statistics, University of Limerick, Limerick, Ireland.\\
d. PepsiCo International, 4 Leycroft Road, Leicester LE4 1ET, United Kingdom
}

\begin{abstract}
%\todo[inline]{Someone needs to write the abstract}
When frying potato snacks, it is typically observed that the dough, which is submerged in hot oil, after some critical time increases its buoyancy and floats to the surface. The lift-off time is a useful metric in ensuring that the snacks are properly cooked. Here we propose a multiphase mathematical model for the frying of potato snacks, where water inside the dough is evaporated from both the top and bottom surfaces of the snack at two receding evaporation fronts. The vapour created at the top of the snack bubbles away to the surface, whereas the vapour released from the bottom surface forms a buoyant blanket layer. By asymptotic analysis, we show that the model simplifies to solving a one-dimensional Stefan problem in the snack coupled to a thin-film equation in the vapour blanket through a non-linear boundary condition. Using our mathematical model, we predict the change in the snack density as a function of time, and investigate how lift-off time depends on the different parameters of the problem.
\end{abstract}

\begin{keyword}

%% keywords here, in the form: keyword \sep keyword

%% MSC codes here, in the form: \MSC code \sep code
%% or \MSC[2008] code \sep code (2000 is the default)

\end{keyword}

\end{frontmatter}

%%
%% Start line numbering here if you want
%%
%% \linenumbers

%% main text
\section{Introduction}
%% Begin Stacie Add
Frying is one of the most common and oldest forms of food cooking. Frying has multiple functions including to sterilise, dehydrate and create product texture~\cite{ModellingFoodProcessingOperations}. Generally there are two types of frying: shallow-fat frying and deep-fat frying. Here we focus on deep fat frying in which the food product being cooked is fully immersed in the oil. 
%% End Stacie Add
During deep-fat frying, some food products undergo density changes that cause them to rise within the oil bath. This process can be exploited in food manufacturing, either as a way of determining the stage of cooking, or as a mechanism to collect the food from the hot oil. For example, in the production of potato snacks, uncooked snacks are submerged in hot oil by a conveyor belt; as the dough cooks they become buoyant and then detach from the conveyor belt.
This must happen at precisely the right moment in order to maximise product quality and the productivity of the process. 
%% Begin Stacie Remove
%% and rise to a second conveyor on which they continue to cook. This
%% is a critical stage in manufacturing since if the uncooked crisps
%% do not detach they become jammed at the end of the conveyor and
%% must be manually removed, which is time consuming and costly.
%% End Stacie Remove
To ensure that the snacks robustly detach at the right time, a better quantitative understanding of the underlying mechanism is needed. 
%In this study we develop a mathematical model describing how the crisp becomes buoyant due 
In particular, there are two major contributions to buoyancy due to the generation of steam, which either escapes from the snack causing a reduction in density, or becomes trapped underneath the snack in a vapour blanket.
%, enhancing buoyancy but retarding heat conduction.

Several different mathematical modelling approaches can be found within the food frying literature. For a comprehensive summary of all relevant types of models, see \cite{moreirabook}. Many of these emphasise transport mechanisms of gases and liquids in porous media \cite{moreirabook, moreira1998new, yamsaengsung2002modeling, chen1997modelling,halder2007improved}.
A commonly used modelling approach is the \textit{crust-core} model, in which there are two regions: a dry \textit{crust} where the water has evaporated, and a wet \textit{core}. In the crust-core approach, mass and energy equations are used to describe the heat and flow in each region, and a moving boundary tracks the evaporation front at the crust-core interface. 
%However, in our case we assume the oil temperature to be constant.
One major concern in the deep-fat frying literature is oil uptake into the snack, and several experiments have been gathered regarding this issue \cite{moreira1997factors, ziaiifar2008review, moyano2006kinetics,bouchon2001oil,kawas2001characterization}. However, most of these models focus on the oil absorbtion post-frying, since this is when most of the oil (approximately 80\%) enters the snack \cite{moreira1998new, yamsaengsung2002modeling}.  Another important objective of many of these studies is to predict quality changes (puffiness, shrinkage, etc) in the snacks as they fry \cite{kawas2001characterization, moreirabook, yamsaengsung2002modeling}. Some models also account for the decrease in the temperature of the oil due to moisture loss from the chip \cite{moreirabook,chen1997modelling}.

A dominant feature of the frying process is the evaporation of the water, which can be observed both from bubbles rising to the surface, and in a vapour layer surrounding the snack. Despite the formation of a vapour blanket being mentioned in several papers (see for instance \cite{halder2007improved} where it is stated that the bubbles impede oil inflow through the bottom boundary) this process has not been described in mathematical terms before within the deep-fat frying literature. In other contexts, film boiling has been studied and expressions for the vapour layer thickness have been derived, as well as bubble generation and release frequencies \cite{dhir1972viscous,board1976recent}. However, none of the above studies address the density changes undergone due to the formation of the vapour blanket, and lift-off is not investigated at all. Furthermore, the effect of the vapour layer, which is a poor conductor, on the heat transfer in the snack is also not discussed.

In this study, we focus on predicting when a snack becomes buoyant, which happens within a few seconds of being introduced into the fryer. Thus, we do not consider structural changes, which occur later on in the frying process; or oil-uptake, which primarily occurs post-frying. We follow the crust-core modelling approach, and we introduce the novel detail of the formation of a vapour layer under the snack. We show that the timescales associated with evaporation indicate that the formation of the vapour blanket is the dominant mechanism for lift-off.
We model the growth of the vapour blanket by coupling a thin film equation to the moving-boundary problem in the snack.
We show that the insulating features of the vapour blanket play an important role in the dynamics of the evaporation fronts.
Whilst all of the models in the above literature are solved numerically by either finite differences or finite volumes, here we combine both numerical and analytical results and compare them together. In particular, we derive closed form solutions for the long-time behaviour of the evaporation fronts and the shape of the vapour blanket, which are useful for the manufacturing process. Furthermore, we show that lift-off times are crucially dependent on the heat transfer properties of the snack.

The remainder of this paper is organised as follows. In Section \ref{sec:model} we introduce the non-dimensional mathematical model for the thermal and flow problems within the snack and vapour blanket. By exploiting the small size of some dimensionless groups, the problem simplifies to solving an energy conservation equation for each region and a thin-film equation for the vapour blanket. A formula that relates the density of the snack to the vapour blanket thickness and the position of the evaporation fronts is also given.  We first solve our model numerically in Section \ref{sec:numerics} using the enthalpy method, and we are able to identify several regimes in the frying process: a heating period, the formation of the vapour blanket, and a regime where the bubble volume is constant. 
Motivated by these numerical results, and considering that the Stefan number of the problem is large, in Section \ref{sec:quasi} we investigate a further simplification to the model, called the quasi-steady limit. In this limit, where the only time-dependence of the system originates from the motion of the evaporation fronts, we obtain analytical solutions that agree well with the numerical results, and provide insight to the frying behaviour.
%The numerical results motivate a further simplification to the model called the quasi-steady limit, which we study in Section \ref{sec:quasi}. In this limit, which is achieved by considering the large Stefan number, . This simplifies the model even further and allows us to obtain analytical solutions that agree well with the numerical results.
We discuss our key findings and their relevance to the snack frying process in Section \ref{sec:conclusion}.

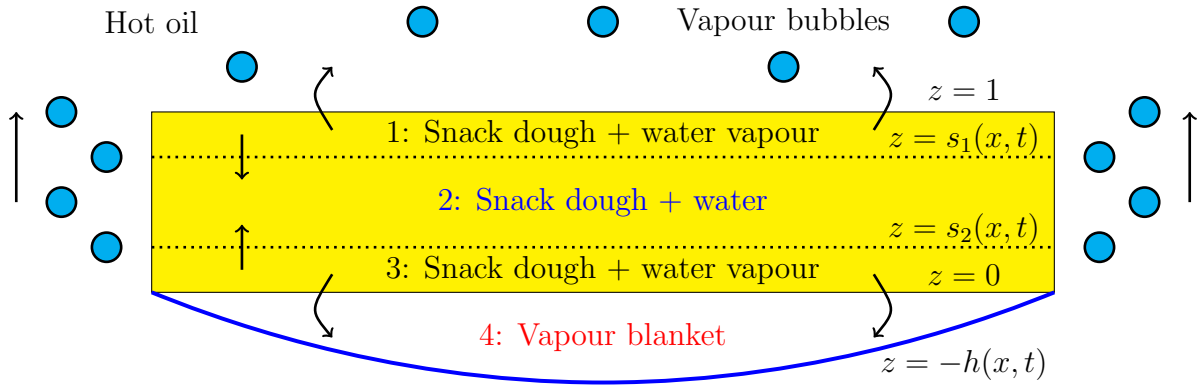
\begin{figure}
\centering
\begin{tikzpicture}[scale=1.2]
\draw [fill=yellow] (0,0) rectangle (10,2);
\draw [line width= 1.5,blue] (0,0) parabola bend (5,-1) (10,0);
\draw [line width= 1,black,dotted] (0,0.5) -- (10,0.5);
\draw [line width= 1,black,dotted] (0,1.5) -- (10,1.5);
%\draw [line width=1, black]  (2.4,0) .. controls (2.1,-0.3) .. (2.2,-0.7);
\node [draw, line width=1, circle, fill = cyan]at (-0.5,0.5){} ;
\node [draw, line width=1, circle, fill = cyan]at (-1,1){} ;
\node [draw, line width=1, circle, fill = cyan]at (-0.5,1.5){} ;
\node [draw, line width=1, circle, fill = cyan]at (-1,2){} ;
\node [draw, line width=1, circle, fill = cyan]at (10.5,0.5){} ;
\node [draw, line width=1, circle, fill = cyan]at (11,1){} ;
\node [draw, line width=1, circle, fill = cyan]at (10.5,1.5){} ;
\node [draw, line width=1, circle, fill = cyan]at (11,2){} ;
\node [draw, line width=1, circle, fill = cyan]at (1,2.5){} ;
\node [draw, line width=1, circle, fill = cyan]at (3,3){} ;
\node [draw, line width=1, circle, fill = cyan]at (5,3){} ;
\node [draw, line width=1, circle, fill = cyan]at (7,2.5){} ;
\node [draw, line width=1, circle, fill = cyan]at (9,3){} ;
\draw [line width=1, black,->]  (2,0.2) .. controls (1.75,-0.2) .. (2,-0.5);
\draw [line width=1, black,->]  (8,0.2) .. controls (8.25,-0.2) .. (8,-0.5);
\draw [line width=1, black,->]  (2,1.8) .. controls (1.75,2.2) .. (2,2.5);
\draw [line width=1, black,->]  (8,1.8) .. controls (8.25,2.2) .. (8,2.5);
\node[red] at (5,-0.5){4: Vapour blanket} ;
\node[blue] at (5,1){2: Snack dough $+$ water} ;
\node[black] at (0,3){Hot oil} ;
\node[black] at (5,1.75){1: Snack dough $+$ water vapour} ;
\node[black] at (5,0.25){3: Snack dough $+$ water vapour} ;
\draw [line width= 1,black,->] (-1.5,1) -- (-1.5,2);
\draw [line width= 1,black,->] (11.5,1) -- (11.5,2);
\draw [line width= 1,black,->] (1,0.25) -- (1,0.75);
\draw [line width= 1,black,->] (1,1.75) -- (1,1.25);
\node at (9,0.7){ $z=s_2(x,t)$} ;
\node at (9,1.7){ $z=s_1(x,t)$} ;
\node at (9,2.25){ $z=1$} ;
\node at (9,0.2){ $z=0$} ;
\node at (9,-0.8){ $z=-h(x,t)$} ;
%\draw [line width= 1,black,<->] (0,-1.3) -- (10,-1.3);
%\node at (5,-1.6){ $L$} ;
%\draw [line width= 1,black,<->] (11.5,0) -- (11.5,2);
%\node at (11.8,1){ \rotatebox{270}{$H$}} ;
\node at (7,3){Vapour bubbles} ;
\end{tikzpicture}
%\end{center}
\caption{ Schematic diagram of the different regions in the snack.}    \label{fig:sketch_crisp} 
\end{figure}

	%\begin{figure}[H]
    %\centering
   % \includegraphics[width=0.8\linewidth]{Crisp_sketch.pdf}
  %  \caption{ Schematic of the different regions in a crisp. }    \label{fig:sketch_crisp} 
%\end{figure}

\section{A Multiphase model for snack frying}\label{sec:model}
In Figure \ref{fig:sketch_crisp}, we illustrate the scenario considered. 
We focus on the two-dimensional case, as shown in the diagram, but we keep the formulation of our mathematical model in three dimensions to be as general as possible.
We propose that the snack is divided into four regions, containing different combinations of dough, water and water vapour. Initially, we assume the dough to be entirely composed of a liquid (water) and solid phase (potato), which is defined as region 2 in our diagram. When the snack is introduced into the fryer, the water begins to evaporate, starting from the exterior. This creates two outer layers containing water vapour and solid, which we denote regions 1 and 3. As the water evaporates from the upper evaporation front, it is bubbled away into the surrounding oil. By contrast, water evaporating from the lower front forms a vapour layer beneath the snack, which we denote region 4.

In this section, we present a non-dimensional mathematical model for the frying of a long thin snack, which consists of energy, mass and momentum conservation equations for each of the different regions of the snack. We simplify these equations by exploiting small parameters in the system. For predicting lift-off time, we introduce a relation between the density of the snack, the size of the vapour blanket and the position of the evaporation fronts.
 
\subsection{Mathematical model}

First, we present the governing equations for each of the regions in Figure \ref{fig:sketch_crisp}.
We keep all the equations in non-dimensional form for convenience, but later we provide further discussion on the derivation, including a list of how each non-dimensional parameter is defined.
As illustrated in the diagram, the domain is long and thin with aspect ratio $\varepsilon=H/L\ll1$. 
We model regions 1 and 3 using an advection-diffusion equation for the temperature, and Darcy's law for the fluid
\begin{align}
\frac{1}{\text{St}}\pdv{T_i}{t}
    + \text{Pe}\left(w_i\pdv{T_i}{z}
    +\varepsilon^2\mathbf{u}_i\cdot\nabla_{xy}T_i\right) &=\pdv[2]{T_i}{z}+\varepsilon^2\nabla_{xy}^2 T_i,&\quad i=1,3,\\
\mathbf{u}_i&=-\nabla P_i,&\quad i=1,3,\\
0&=\pdv[2]{P_i}{z}+\varepsilon^2\nabla_{xy}P_i,&\quad i=1,3,
\end{align}
where $T_i(\mathbf{x}_i,t)$ is the temperature, $\mathbf{u}_i(\mathbf{x},t) = (u_i,v_i,w_i)$ is the velocity of the fluid, and $P_i(\mathbf{x},t)$ is the pressure. Subscripts are used to denote the different regions and $\nabla_{xy} = (\frac{\partial }{\partial x},\frac{\partial }{\partial y},0)$ is the gradient in the $x$-$y$ plane. Our dimensionless parameters are the P\'eclet number $\text{Pe}$, and the Stefan number $\text{St}$. We assume that the flow in the core region 2 of the snack is negligible, and so there is no need for any mass or momentum equations. The heat equation in this region is
 \begin{align}
 \frac{\text{C}}{\text{St}}\pdv{T_2}{t}=\text{K}_1\left(\pdv[2]{T_2}{z}+\varepsilon^2\nabla_{xy}^2 T_2\right),
 \end{align}
where $\text{K}_1$ and $\text{C}$ are the ratios of thermal conductivities and volumetric heat capacities between regions 2 and 1. In region 4, we have an advection-diffusion equation for the temperature and the Navier-Stokes equations for the fluid flow,
 \begin{align}
 \dfrac{\text{Pe}}{\text{K}_2}\left(\frac{1}{\tau}\pdv{T_4}{t}
+\textbf{u}_4\cdot \nabla T_4\right)&=\pdv[2]{T_4}{z}+\varepsilon^2\nabla_{xy}^2 T_4, \\
%Momentum
\text{Re}\left(\frac{1}{\tau}\,\pdv{\mathbf{u}_4}{t}
+ \mathbf{u}_4\cdot \nabla \mathbf{u}_4\right)
&=
-\beta \left[\pdv{P_4}{x},\pdv{P_4}{y},\dfrac{1}{\varepsilon^2}\pdv{P_4}{z} \right]^{\text{T}}
+ \pdv[2]{\mathbf{u}_4}{z}+\varepsilon^2\nabla_{xy}^2\mathbf{u}_4-\frac{\text{Re}}{\text{Fr}^2}\hat{\mathbf{z}},\label{NS4}\\
% Incompressubility
\nabla \cdot \mathbf{u}_4&=0, \label{continuity}
 \end{align}
where $\text{K}_2$ is a ratio of thermal conductivities between regions 4 and 1, $\tau$ is the ratio of the timescale of evaporation to the timescale of evolution of the vapour blanket $z=-h$, 
$\text{Re}$ is the Reynolds number, $\text{Fr}$ is the Froude number, and $\beta$  is a measure of the relative size of the hydrostatic pressure of the oil acting on the gas in region 4 and the pressure drop needed to maintain the Darcy gas flux in regions 1 and 3. On the boundaries at $z=1$ and $z=-h$, we have Newton's law of heating
 \begin{align}
 % Oil dry crisp interface
\frac{1}{\text{N}}\pdv{T_1}{z}&=1-T_{1}, \quad &z&=1\\
\dfrac{\text{K}_2}{\text{N}\sqrt{1+\varepsilon^2 (\nabla_{xy} h)^2}}
\left( 
\pdv{T_4}{z}+\varepsilon^2\nabla_{xy} h \cdot \nabla_{xy} T_4
\right)&=T_4-1,
 &z&=-h,
\end{align}
where $\text{N}$ is the Nusselt number, measuring the ratio between heat transfer at the boundary and heat conduction in the snack. At the boundary, $z=-h$, we have the kinematic and dynamic boundary conditions
\begin{align}
 \frac{1}{\tau}\pdv{h}{t}&=w_4-\mathbf{u}_4\cdot\nabla_{xy} h,  &z&=-h,\label{flowcoup1}\\
\textbf{D} \cdot \textbf{n}&
=\dfrac{\beta}{\varepsilon}\left( P_4-h-\dfrac{\kappa}{\text{Bo}}\right)\textbf{n},
&z&=-h,\label{flowcoup2}
\end{align}
where $\text{Bo}$ is the Bond number, $\kappa$ is the curvature, $D$ is the strain rate tensor, and $\mathbf{n}$ is the normal to the vapour blanket, given by
 \begin{align}
 \kappa&=\dfrac{\nabla_{xy}^2 h}{(1+\varepsilon^2(\nabla_{xy} h)^2)^{3/2}},\\
\mathbf{D}&=
\begin{pmatrix}
2\varepsilon\pdv{u_4}{x}                    & \varepsilon^2(\pdv{v_4}{x}+\pdv{u_4}{y}) & \varepsilon^2\pdv{w_4}{x}+\pdv{u_4}{z}\\
\varepsilon^2(\pdv{u_4}{y}+\pdv{v_4}{x}) & 2\varepsilon\pdv{v_4}{y}                    & \varepsilon^2\pdv{w_4}{y}+\pdv{v_4}{z}\\
\pdv{u_4}{z}+\varepsilon^2\pdv{w_4}{x}   & \pdv{v_4}{z}+\varepsilon^2\pdv{w_4}{y}   & 2\varepsilon\pdv{w_4}{z} \\
\end{pmatrix}\\
\mathbf{n}&=
\dfrac{1}{\sqrt{1+\varepsilon^2(\nabla h)^2}}
\begin{pmatrix}
-\varepsilon\pdv{h}{x}\\
-\varepsilon\pdv{h}{y}\\
1
\end{pmatrix}.
\end{align}
On the evaporation fronts, $z=s_i$ for $i=1,2$, we require that the temperature matches the evaporation temperature of water
\begin{align}
 T_1&=T_2=0,   &z&=s_1,\\
 T_2&=T_3=0, &  z&=s_2.
\end{align}
We also have a Stefan condition describing the motion of the evaporation fronts. This condition can be derived by balancing the latent energy required to vaporise water with difference in heat flux on either side of the boundary. This gives us
\begin{align}
\dot{s}_1 &= \text{K}_1 \pdv{T_2}{z} - \pdv{T_1}{z}
+ \varepsilon^2 \nabla_{xy} s_1 \cdot \nabla_{xy} (T_1 - \text{K}_1 T_2), &z &=s_1,\\
\dot{s}_2 &= \text{K}_1 \pdv{T_2}{z} - \pdv{T_3}{z}
 + \varepsilon^2 \nabla_{xy} s_2 \cdot \nabla_{xy} (T_3 - \text{K}_1 T_2), &z &= s_2.
\end{align}
The change in density undergone when the water vaporises creates a flow in regions 1 and 3. As discussed by \cite{myers2019stefan}, the equations that describe this volume change generated flow are
\begin{align}
\lb1-\frac{1}{\text{R}}\rb\dot{s}_1&=-w_1+\varepsilon^2\mathbf{u}_1\cdot\nabla_{xy} s_1,
&z&= s_1,\label{flowcoup3}\\
\lb1-\frac{1}{\text{R}}\rb\dot{s}_2&=-w_3+\varepsilon^2\mathbf{u}_3\cdot\nabla_{xy} s_2,
&z&= s_2,\label{flowcoup4}
\end{align}
where R is the ratio of the density of water to the density of steam. This signifies the volume change that happens when the water is vaporised, which drives the gas flow.
  
Finally, at the interface between the snack and the vapour blanket, we have continuity of temperature, mass, pressure, and heat flux
   \begin{align}
% Dry crisp / Gas pocket
 &T_3=T_4, \quad P_3=\varepsilon^{-2} \beta \Gamma P_4,\quad \text{K}_2 \pdv{T_4}{z}=\pdv{T_3}{z},\quad &z&=0,\label{conditionsT4} \\
&\left[ \varepsilon^2 u_3,\varepsilon^2v_3,w_3\right]
=\left[ u_4, v_4,w_4\right]  &z&=0,
% Gas pocket oil
 \end{align}
 where $\Gamma$ is the non-dimensional permeability of the snack.
The dimensionless initial conditions are given by 
\begin{align}
T_2(\textbf{x},t)&=T^*\label{nondimic1},\\
s_1(x,y,0)&=1,\label{nondimic2}\\
s_2(x,y,0)&=0,\label{nondimic3}\\
h(x,y,0)&=0.\label{nondimic4}
\end{align}
 In Table \ref{dimensionless_parameters} we list the dimensionless parameters of the system, their definitions in terms of dimensional parameters, and their approximate values.
\begin{table}[h]
\centering
\begin{tabular}{c c c c}
      \hline 
      Parameter  & Definition & Value  \\ 
      \hline
     $\text{St}$ & ${L_v\alpha \rho_l} / ({\rho_1 c_{p,1}(T_o-T_{e})})$ &  $8.4$\\
      $\text{C}$ & ${c_{p,2}\rho_2}/({c_{p,1}\rho_1})$ &$2.1$\\
      $ \varepsilon$ & $H/L$& $ 1.1 \times 10^{-2}$ \\
      $ \text{Pe}$&$ {c_{p,4}(T_o-T_{e})}/{L_v}$ &$5.8\times 10^{-2}$\\
      $\text{K}_1$&${k_2}/{k_1}$&$ 1.4$\\
      $\text{K}_2$ & ${k_4}/{k_1}$&$4.2 \times 10^{-2}$\\
      $\tau$  &${\alpha \rho_l}/{\rho_v}$ & $5.8\times 10^{2}$ \\
      $ \text{Re}$&${k_1 (T_o-T_{e})}/({L_v \mu_v})$&$9.2 \times 10^{-1}$\\
      $\beta$&${L_v g \rho_v\rho_o H^5}/({k_1L^2\mu_v(T_o-T_{e})})$&$6.5\times 10^{-1}$\\
      $\Gamma$&$\chi/H^2$&$2.0 \times 10^{-4}.$ \\
      $\text{N}$&${h_c H}/{k_1}$&$ 1.3$\\
      $T^*$&$({T_{\text{a}}-T_{e}})/({T_o-T_{e}})$&$-1.1$\\
      $\text{Bo}$&$\rho_o g L^2/\gamma$&$1.1\times 10^3$\\
       $\text{R}$ & $\rho_l/\rho_v$ & $1.7 \times 10^{3}$\\
       $\text{Fr}$ & $(k_1(T_o-T_e)/\rho_v L_v H^2) \sqrt{L/g}$ & $4.3$\\
      \hline
    \end{tabular}
    \caption{Dimensionless parameters and their approximate numerical values.}
    \label{dimensionless_parameters}
\end{table}
The dimensional parameters appearing in Table \ref{dimensionless_parameters} are: $L_v$, the latent heat of vaporisation of water; $\alpha$, the porosity of the snack; $\chi$, the permeability of the snack; $h_c$, the heat transfer coefficient; $\gamma$, the interfacial tension between water and oil; $g$, the acceleration due to gravity; $H$, the height of the snack; $L$, the length of the snack; $\mu_v$, the viscosity of water vapour; $T_o$, $T_l$, $T_{\text{a}}$, the temperature of the oil, the evaporation temperature of water, and the ambient air temperature of the snack before entry into the oil; and $\rho_o$, $\rho_l$, $\rho_v$, the density of oil, water, and vapour, respectively. Finally, we have some compound parameters for regions 1, 2, and 3, each with a subscript denoting the relevant region. These are: $\rho_j$, the compound density; $k_j$ the compound thermal conductivity; and $c_{p,j}$, the compound specific heat capacity, with $j=1-4$. These compound parameters have been determined by taking a volume-weighted-average of the parameters for the individual phases (solid snack, water, vapour) in each region (see for instance, \cite{kiradjiev2019}). For example, $\rho_1=\alpha_s\rho_s+\alpha_v\rho_v$, where each $\alpha$ represents a mass fraction.

\subsection{Model simplifications}
Having calculated the non-dimensional parameters in Table \ref{dimensionless_parameters}, we are motivated to consider the asymptotic limit of
\beq
\text{Pe},\,\varepsilon,\,\text{Re},\,\text{Bo}^{-1},\,\text{Pe}/\text{K}_2,\,\text{Pe}/\text{K}_2\tau,\,\text{Re}/\tau,\,\varepsilon^2\text{K}_1,\,\text{Re}/\text{Fr}^2,\,1/\text{R} \rightarrow 0.
\eeq
Note that although a few of these parameter groups associated with the vapour layer (Re, Pe/K$_2$) are marginal in this scaling, we have also carried out a more complex scaling in which the thicknesses of regions 3 and 4 are scaled separately. This scaling confirms that all the dimensionless quantities listed above are small. 

Under these limits, the only coupling between the flow and thermal problems is through the boundary conditions \eqref{flowcoup1}, \eqref{flowcoup2}, \eqref{flowcoup3} and \eqref{flowcoup4}. The simplified governing equations for the heat problem are
\begin{align}
\frac{1}{\text{St}} \pdv{T_1}{t}&=\pdv[2]{T_1}{z},&\quad s_1\leq z\leq 1 \label{region1eq},\\
\frac{\text{C}}{\text{St}} \pdv{T_2}{t}&=\text{K}_1\pdv[2]{T_2}{z},&\quad s_2\leq z\leq s_1\label{region2eq},\\
\frac{1}{\text{St}} \pdv{T_3}{t}&=\pdv[2]{T_3}{z},&\quad 0\leq z\leq s_2\label{region3eq}.
\end{align}
The only dependence of the heat problem on the thickness of the vapour blanket $h$ is through the lower boundary condition. Hence, the complete set of boundary conditions for the heat problem are
\begin{align}
% Oil dry snack interface
\frac{1}{\text{N}}\pdv{T_1}{z}   &=1-T_1,  &z&=1,\label{nondimbc0}\\
% Dry/Wet crisp
T_1&=T_2=0,   &z&= s_1,\label{nondimbc1}\\
\dot{s}_1&=\text{K}_1\pdv{T_2}{z}-\pdv{T_1}{z}, &z &= s_1,\label{nondimbc2}\\
%\dot{s}_1&=w_1 &z&=s_1,\\
% Wet dry crisp
T_2&=T_3=0,  &z&=s_2,\label{nondimbc3}\\
\dot{s}_2&=\text{K}_1\pdv{T_2}{z}-\pdv{T_3}{z}, &z&=s_2,\label{nondimbc4}\\
%\dot{s}_2&=-w_3 &z&= s_2,\\
% Dry crisp / Gas pocket
\frac{1}{\text{N}}\pdv{T_3}{z}\left(\frac{h\text{N}}{\text{K}_2}+1\right)&=T_3-1, &z&=0,\label{nondimbc5}
\end{align}
where equation \eqref{nondimbc5} is derived by solving for $T_4$ and inserting the solution into \eqref{conditionsT4}.
Specifically, $T_4$ is given in terms of $T_3$ and $h$ by
\begin{equation}
	T_4=\frac{1}{\text{N}}\left.\pdv{T_3}{z}\right|_{z=0} \left(\frac{(z+h)\text{N}}{\text{K}_2}+1\right)+1.
\end{equation}
In order to obtain an equation for $h$, we need to follow a series of steps.
Firstly, taking the third component of the simplified version of \eqref{NS4} together with the reduced form of \eqref{flowcoup2} we obtain $P_4=h$ throughout region $4$. Now, 
$\mathbf{u}_4$ can be found simply by integrating the reduced form of the first two components of \eqref{NS4}, as well as \eqref{continuity}. Substituting this into the kinematic condition \eqref{flowcoup1} gives 
\begin{equation}
\frac{1}{\tau}\pdv{h}{t}=\dfrac{\beta}{3}\nabla_{xy}\cdot\left( h^3\nabla_{xy} h \right)-w_4|_{z=0}.
\end{equation}
Finally, by considering the fluid problem in region 3, and using the simplified version of \eqref{flowcoup4}, we see that $w_4|_{z=0} = \dot{s}_2$. Thus, the governing thin-film equation for the vapour blanket becomes
\begin{equation}
\frac{1}{\tau}\pdv{h}{t}=\dfrac{\beta}{3}\nabla\cdot\left( h^3\nabla h \right)+\dot{s}_2.\label{region4eq}
\end{equation}
%Note that the quantity $(1-\text{R})/\tau\approx2.9$ is considered neither large nor small, so we keep the term on the right hand side of (\ref{region4eq}).
We would expect that at the edges of the snack, $h$ would take some finite value and the pressure would be equivalent to the hydrostatic pressure of the oil. However, in our thin film equation (\ref{region4eq}) we cannot impose both conditions, so we choose
\beq
h=0, \text{ at $\delta \Omega_0$}, \label{flowbcs}
\eeq
as the lateral boundary condition, where $\Omega_0$ is the cross-section of the snack at $z=0$, and $\delta \Omega_0$ is the boundary of $\Omega_0$.

\begin{comment}

As equation~\ref{region4Teq} shows that the gas layer is always in thermal equilibrium we can propagate the thermal boundary condition at $z=-h$ to the $z=0$ surface of the crisp. Equations~\ref{region4Teq} and~\ref{nondimbc5} give $\pdv{T_4}{z}=\left[T_4(-h)-T_3(0)\right]/h$. Combining this equation with equations~\ref{nondimbc55} and~\ref{nondimbc6} we can eliminate $T_4$ and its gradient to incorporate the effects of the gas layer by a heat transfer boundary condition at $z=0$. Thus we have
\begin{equation}
    \left(\dfrac{1}{\text{N}}+\dfrac{h}{\text{K}_2} \right)\pdv{T_3}{z}=T_3-1, \quad z=0.
\end{equation}
Physically we can interpret this as two thermal resistances in series, one corresponding to heat transfer at the oil-gas interface and the other corresponding to the heat transfer through the gas layer. 

\end{comment}

Note that whilst the vapour blanket thickness depends spatially on $x$ and $y$, $h=h(x,y,t)$, the temperature only depends on $z$, except for the boundary condition (\ref{nondimbc5}). Hence, it is convenient to replace $h$ in (\ref{nondimbc5}) by an average film thickness $\bar{h}=\int_{\partial \Omega_0} h \, dx dy$. Making this substitution, the thermal problem is purely in terms of $z$, and the vapour blanket problem is in terms of $x$ and $y$. We can simplify even further by assuming that the snack is uniform in the $y$ direction, giving us a one-dimensional model for the thermal problem in $z$, and a one-dimensional model for the vapour blanket problem in $x$. This is the approach that we take for the remaining of the paper.

%\subsection{Further Simplifications}
%These equations can be further simplified. In region 1 the equations relating to $\mathbf{v}_1$ and $P_1$ decouple from the other equations and so it is unnecessary to solve these equations. 

\begin{comment}
Finally we can replace the heat equations in regions 1, 2 and 3 by an enthalpy
equation.
\begin{equation}
    \pdv{\Phi}{t}=\pdv[2]{T}{z}
\end{equation}
where the dimensionless temperature $T$ and dimensionless
enthalpy, $\Phi$, are related by
\begin{equation}
    T=
    \begin{cases}
    \text{St}_2\Phi & \Phi<0\\
    0 & 0\leq \Phi\leq 1\\
    \text{St}(\Phi-1)
    & \Phi>1
    \end{cases}
\end{equation}
\end{comment}

\subsection{Density calculation and lift-off time}

A necessary condition for the snack to detach from the conveyor belt is that its density is less than that of the surrounding oil. The reduction of the density of the snack is due to two processes. Firstly there is loss of mass as water evaporates into steam and leaves the snack. Secondly the formation of the vapour blanket increases the volume of the snack. 

%In dimensional terms, as a function of time, the density of the crisp is given by 
%\begin{equation}
%\rho_{\text{crisp}}(t)=\frac{1}{\mathrm{Vol}(t)}\lb \rho_v \int^{L}_{0} h \, \mathrm{d}x  W %+   \rho_l \alpha_l (s_2-s_1)L W +\rho_v \alpha_v (H-(s_2-s_1))L W + \rho_s \alpha_s L W H %\rb,
%\end{equation}
%where $W$ is the crisp `width' in the third dimension and the volume is given by
%\beq
%\mathrm{Vol}(t)=L W H +  \int^{L}_{0} h \, \mathrm{d}x W.
%\eeq
%\section{Non-dimensional problem}
%We choose the following scalings for the variables
%\beq
%\begin{split}
%x=\ell\hat{x},\quad z=D\hat{z},&\quad T=T_{e}+(T_o-T_{e})\hat{T}, \quad t=\frac{\alpha_l \rho_l L D^2}{k_{sv}(T_o-T_{e})}\hat{t},\\
%&\quad h=D\hat{h},\quad s_i=D\hat{s}_i.
%\end{split}
%\eeq
%We have introduced the following non-dimensional numbers:
%\beq
%\begin{split}
%&\mathrm{St}=\frac{\alpha_l \rho_l L \kappa_{sv}}{k_{sv}(T_o-T_{e})},\quad  \kappa=\frac{\kappa_{sl}}{\kappa_{sv}},\quad K=\frac{k_{sl}}{k_{sv}},\\
%\phi_1=\frac{\alpha_l  \rho_l}{\rho_v} ,&\quad \phi_2 = \frac{\rho_o g D^3 \tau}{\ell^2 \mu_v}, \quad N= \frac{h_{c} D}{k_{sv}},\quad \hat{\delta}=\frac{\delta}{D},\quad \hat{T}^*=\frac{T_{room}-T_e}{T_o-T_e},
%\end{split}
%\eeq
%where $\tau$ is the dimensional timescale.
The dimensionless density, ${\rho}$, is scaled by the density of oil, $\rho_o$, so that ${\rho}=1$ when the snack is neutrally buoyant. The density is given by
\beq
{{\rho}}_{\text{snack}}({t})= \frac{1}{1+\int^{1}_{0} {h} \, \mathrm{d}{x}}\lb \dfrac{\rho_v}{\rho_o} \int^{1}_{0} {h} \, \mathrm{d}{x} +   \dfrac{\rho_l}{\rho_o} \alpha_l ({s}_1-{s}_2)+\dfrac{\rho_v}{\rho_o} \alpha_v \left[1-({s}_1-{s}_2)\right] + \dfrac{\rho_s}{\rho_o} \alpha_s  \rb.
\eeq
The denominator is the volume of the snack, including the volume of the bubble given by integrating over $h$. The numerator is the mass of the snack broken into contributions from the gas in the bubble, liquid water in region 2, water vapour in regions 1 and 3 and the solid component of the snack. 
Therefore, the non-dimensional lift-off time, which we denote $t^*$, is the first time\footnote{Note that in reality, there may be some surface tension effects holding the snack down to the solid substrate, therefore delaying lift-off time. However, since these depend on the specific surface properties of the fryer substrate, we do not study such effects here. } for which
\beq
{\rho}_{\text{snack}}(t^*)<1.\label{oildrop}
\eeq

\section{Numerical Approach} \label{sec:numerics}
%\textbf{Graham}

\begin{figure}
\centering
%\begin{overpic}[width=0.3\textwidth]{T}
%\put(52,12){\color{black}\vector(0,1){50}};
%\put(30,65){Increasing $t$}
%\end{overpic}
%\begin{overpic}[width=0.3\textwidth]{h}
%\put(50,57){\color{black}\vector(0,-1){12}};
%\put(20,37){Increasing $t$}
%\put(65,61){\small Crisp}
%\put(65,43){\small Film}
%\end{overpic}
%\begin{overpic}[width=0.3\textwidth]{rho}
%\end{overpic}
\begin{tikzpicture}[scale=0.5]
\node at (0,0) {\includegraphics[width=0.45\textwidth]{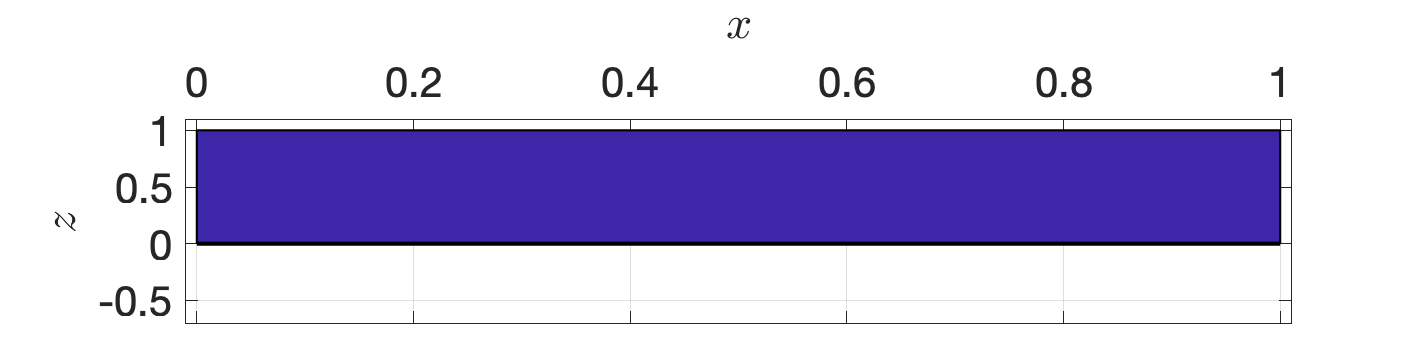}};
\node at (4,2) { $\boldsymbol{{t}=0}$};
\node at (17,0) {\includegraphics[width=0.45\textwidth]{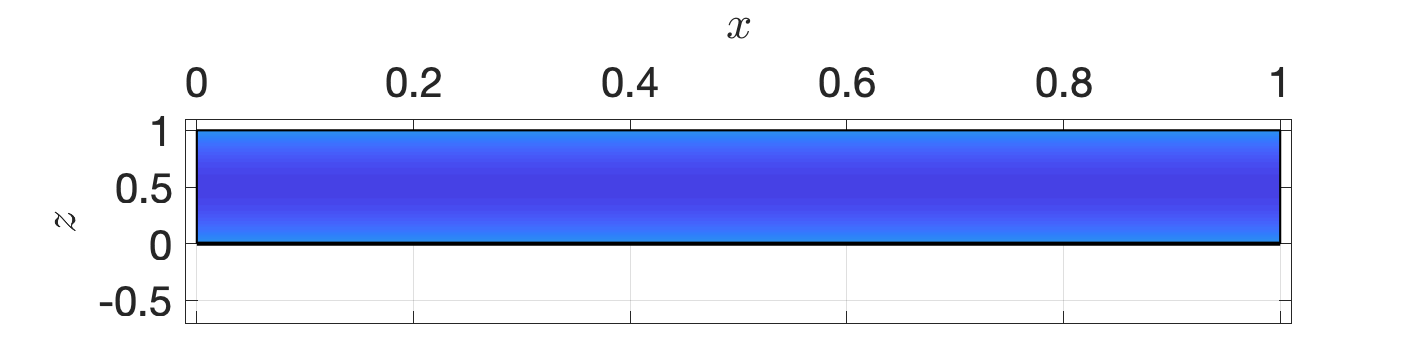}};
\node at (8.5,-0.7) {\includegraphics[width=0.1\textwidth]{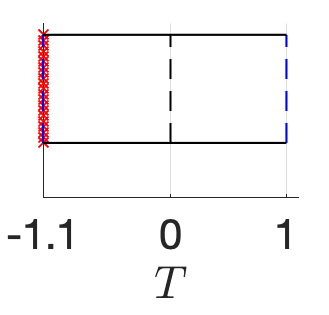}};
\node at (25.5,-0.7) {\includegraphics[width=0.1\textwidth]{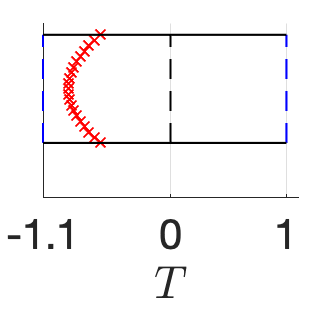}};
\node at (22,2) { $\boldsymbol{{t}=0.02}$};
\node at (0,-4) {\includegraphics[width=0.45\textwidth]{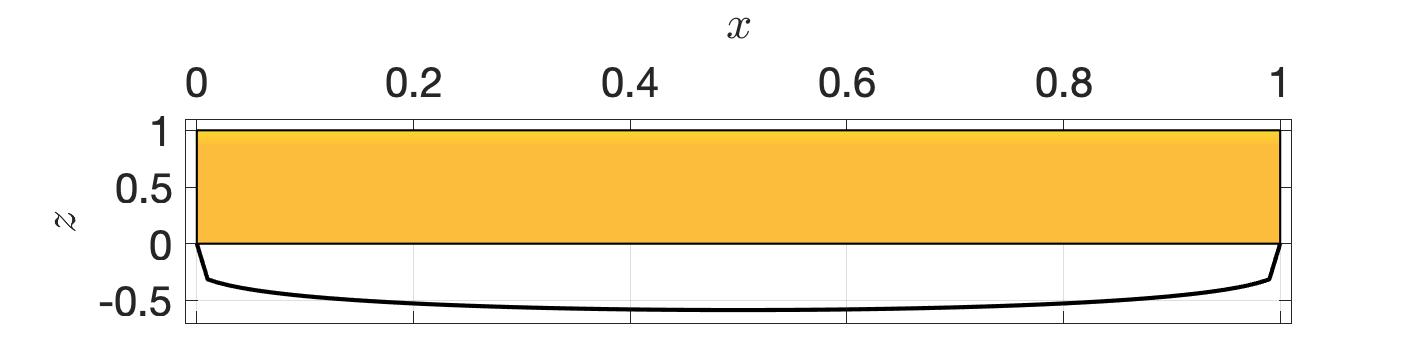}};
\node at (4,-2.1) { $\boldsymbol{{t}=0.3}$};
\node at (17,-4) {\includegraphics[width=0.45\textwidth]{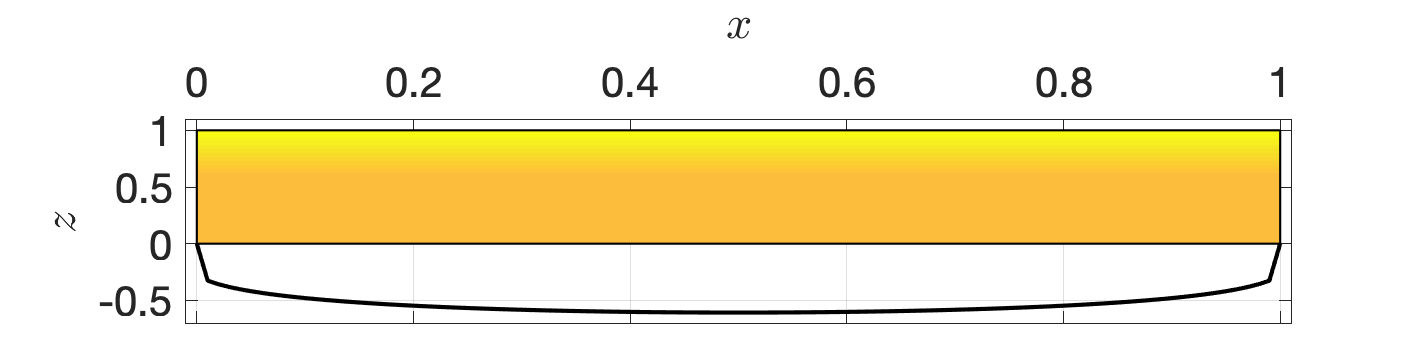}};
\node at (8.5,-4.7) {\includegraphics[width=0.1\textwidth]{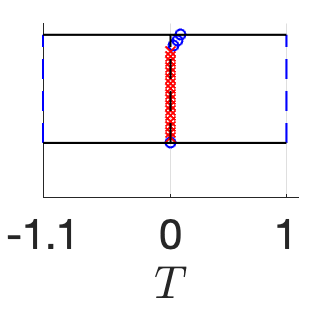}};
\node at (25.5,-4.7) {\includegraphics[width=0.1\textwidth]{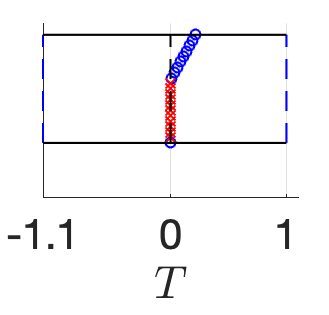}};
\node at (22,-2.1) { $\boldsymbol{{t}=0.8}$};
\node at (7,-8) {\includegraphics[width=0.5\textwidth]{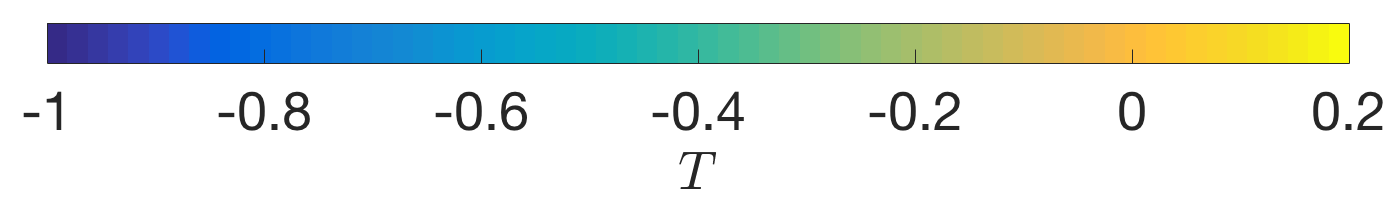}};
\node at (22,-7.5) {\includegraphics[width=0.25\textwidth]{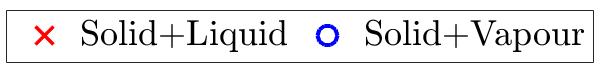}};
\node at (-7,2) {a)}; 
\node at (10,2) {b)}; 
\node at (-7,-2.5) {c)}; 
\node at (10,-2.5) {d)}; 
\end{tikzpicture}
%\vspace{0.5cm}
\begin{tikzpicture}[scale=0.5]
\node at (0,0) {\includegraphics[width=0.6\textwidth]{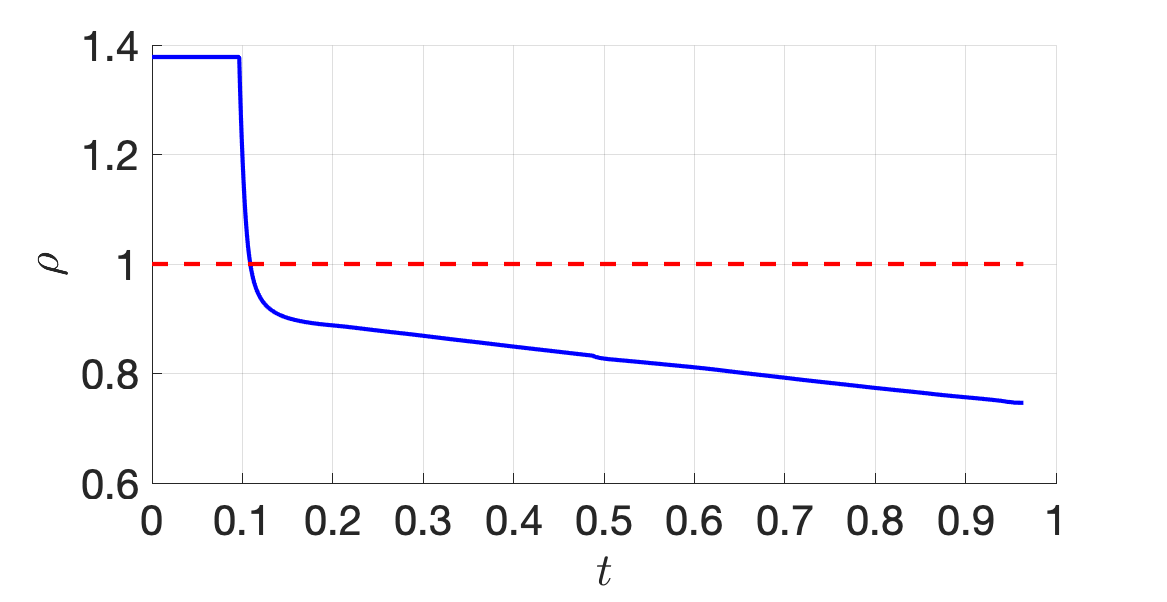}};
\draw[line width=1,<->] (-7.7,5) -- (-6.3,5);
\draw[line width=1,<->] (-6.2,5) -- (-4.9,5);
\draw[line width=1,<->] (-4.8,5) -- (8,5);
\node at (-7.1,-0.2) {\footnotesize \rotatebox{90}{Heat diffusion}};
\node at (-5.3,-0.1) {\footnotesize \rotatebox{90}{Bubble inflation}};
\node at (1,4) {\footnotesize Quasi-steady};
\node at (-10,6) {e)}; 
%\node at (-7,6) {\large A};
%\node at (-6,6) {\large B};
%\node at (1,6) {\large C};
\node at (16,0) {\includegraphics[width=0.37\textwidth]{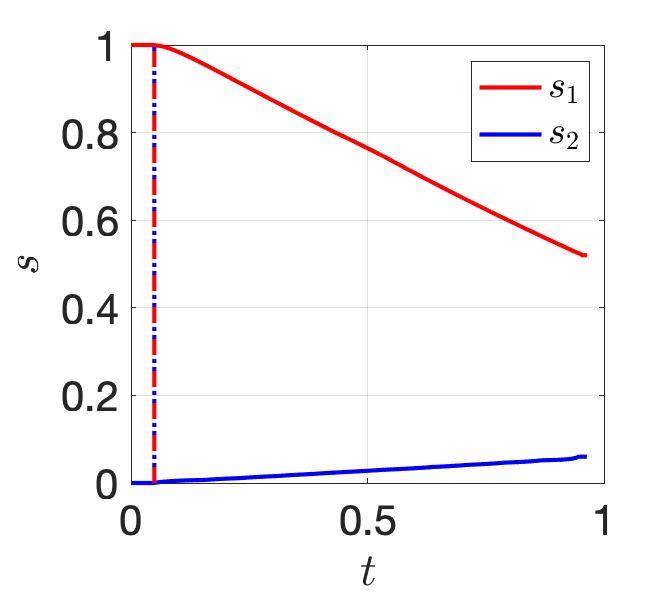}};
\node at (10,6) {f)}; 
\end{tikzpicture}
\caption{(a,b,c,d) Numerical solution at ${t}=0,0.02,0.1,0.8$, showing a colour plot of the temperature in the snack, a corresponding line plot of the temperature, and the film thickness beneath the snack. (e) Density evolution over time, indicating the critical density for lift-off $\rho=\rho_{oil}$. (f) Evolution of the evaporation fronts $s_1$ and $s_2$.
\label{temperature_profile} \label{soln}}
\end{figure}
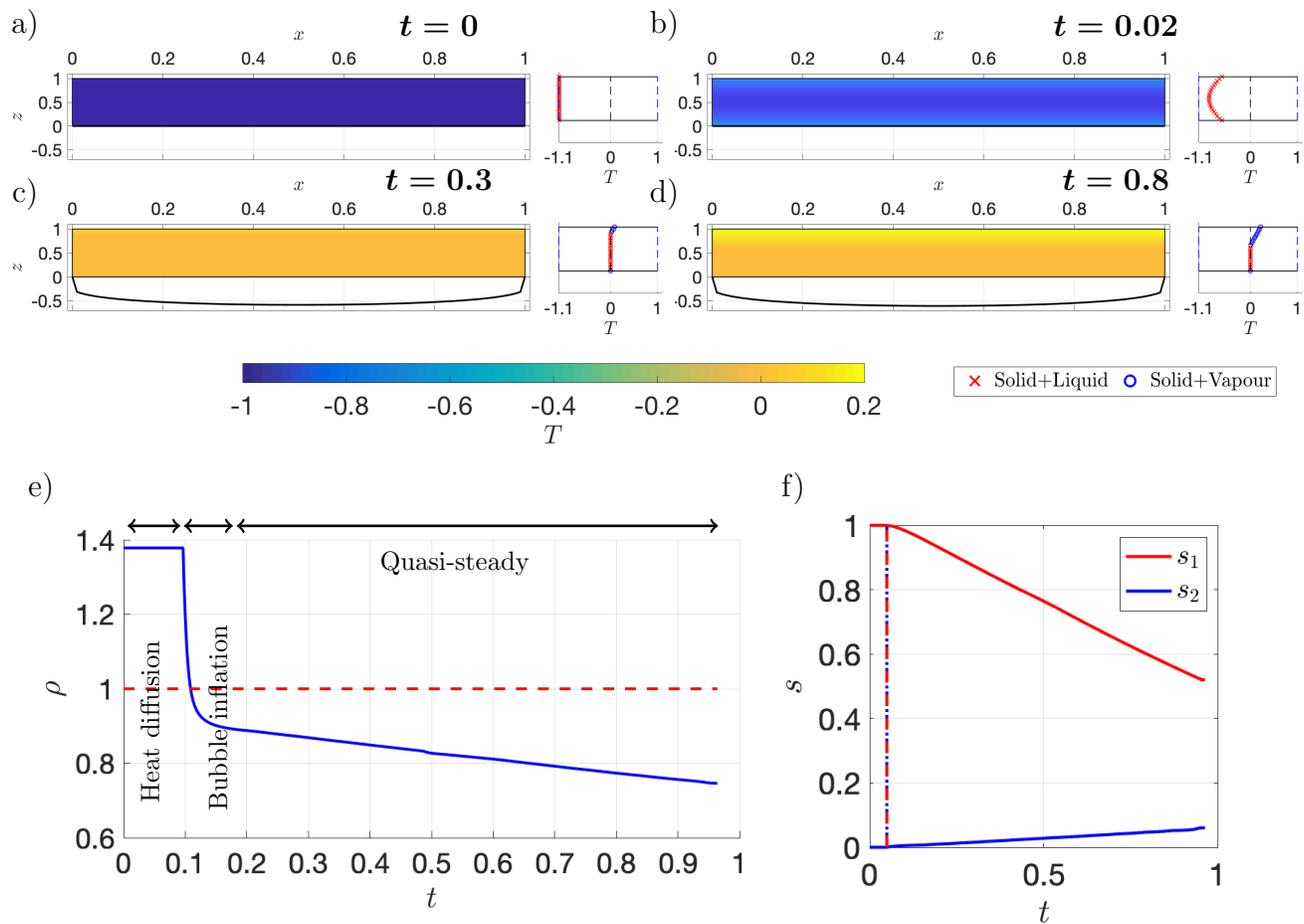

%\subsection{Solution method}

Our first approach is to solve the problem (\ref{region1eq})-(\ref{region3eq}), (\ref{region4eq}) numerically using the enthalpy method \cite{benham2016solidification,voller1981accurate}. 
The non-dimensional temperature is related to the non-dimensional enthalpy in the following way:
\beq
{T}=\begin{cases}
\frac{\mathrm{St}\text{K}_1}{C}{\theta}:&\quad {\theta}<0, \\
0:&\quad 0\leq {\theta} \leq 1, \\
\mathrm{St}{\lb \theta - 1 \rb}:&\quad {\theta}> 1.
\end{cases}\label{entheq}
\eeq
The enthalpy method conveniently reduces the problem to solving the single partial differential equation
\beq
\pdv{{\theta}}{{t}}=\pdv{^2 {T}}{{z}^2},\label{reducedpde}
\eeq
within the entire domain $0\leq{z}\leq 1$, where ${\theta}$ and ${T}$ are related via (\ref{entheq}). 
%The non-dimensional enthalpy is given in terms of the dimensional enthalpy by 
%\beq
%{\theta}=\frac{k_{1}\theta}{\alpha_l \rho_l L_v k_{2}} - \frac{k_{1}\rho_2 %c_{p_2} T_{e}}{\alpha_l \rho_l k_{2} L_v}. 
%\eeq
%Therefore, a change in ${\theta}$ by an amount $1/(\alpha_l  K)$ corresponds to an increase in enthalpy by $\rho_l L_v$, which is the thermal content associated with latent heat. 
We use the method of lines with an explicit forward Euler scheme to solve (\ref{region4eq}) and (\ref{reducedpde}), where at each time step we update ${T}$ using the relation (\ref{entheq}).

%\subsection{Results}

We plot the solution in Figure \ref{soln}, illustrating the evolution of both the temperature, the vapour blanket, and the resultant snack density. We identify several clear regimes in the frying process, which we indicate in the density plot. Initially the snack is plunged into the oil at room temperature, and so the first regime consists of a heating period, bringing the temperature within the snack to the evaporation temperature. During this regime the snack is entirely composed of liquid and solid (region 2). Once the temperature is near the boiling point everywhere, and equal to the boiling temperature at the edges of the snack, the latent heat begins to be removed. As the latent heat is removed from the edges of the snack, two evaporation fronts recede into the interior of the snack, bubbling away vapour through the top and bottom. This is the second regime of the process, during which the vapour blanket is formed, and inflates very rapidly, causing a sudden drop in density. The vapour blanket quickly reaches a steady state, bringing us to the final regime. During this regime, the evaporation fronts continue to move inwards (hence it is called the quasi-steady regime), and the temperature within each region is approximately linear with $z$, which is due to the large Stefan number \cite{benham2016solidification, louro1986remarks, barry2008exact}. Meanwhile the bubble remains at near-constant volume, which can only be explained by a constant growth rate of the evaporation front $\dot{s}_2$ in (\ref{region4eq}).

\begin{figure}
\centering
\includegraphics[width=0.6\textwidth]{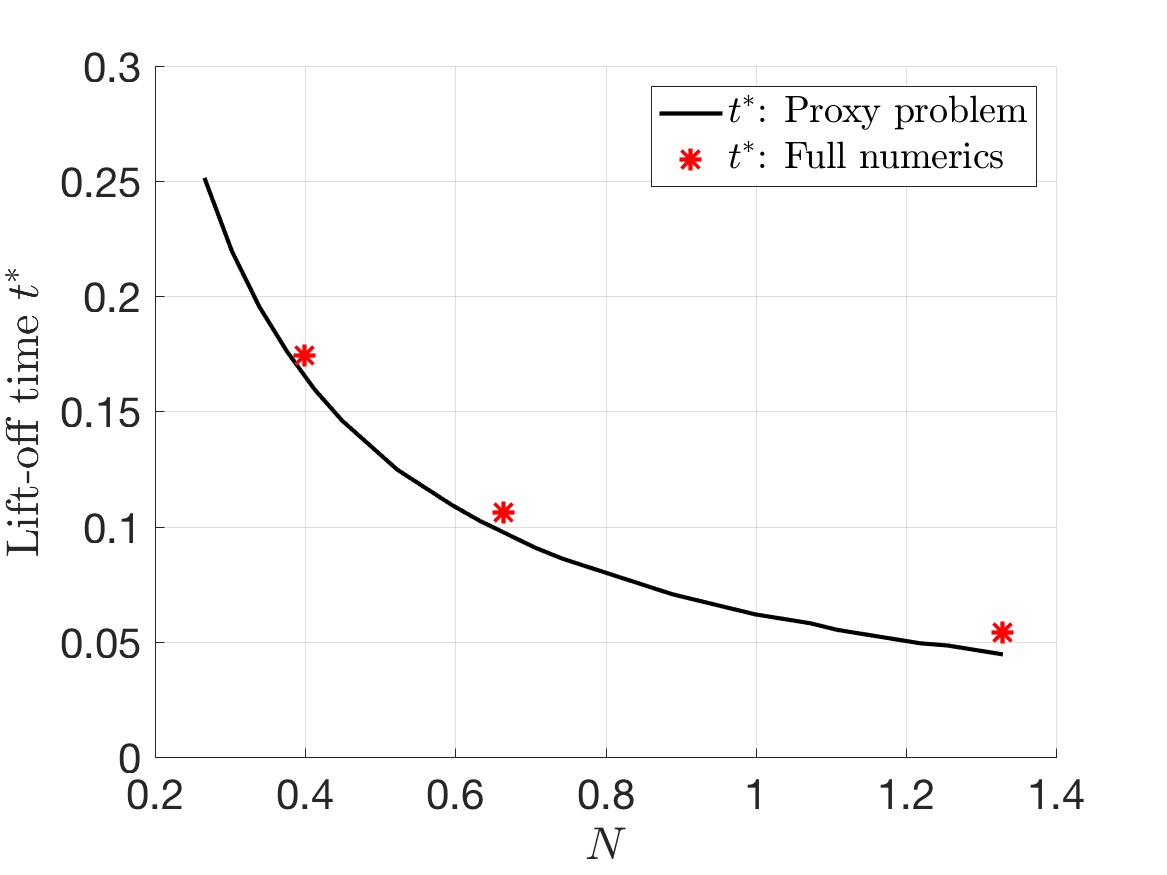}
\caption{Variation of the non-dimensional lift-off time with Nusselt number, showing approximate time calculated by solving the proxy heat diffusion problem until the snack reaches evaporation temperature, and the precise times calculated by solving the full numerical problem until $\rho(t^*)<1$.
\label{nusselt}}
\end{figure}

The lift-off time of the snack can be taken as the time at which the density falls below the oil density (\ref{oildrop}). For the parameters used here, this corresponds to a time of $t=0.1$, or in dimensional terms, $1$ second, which is in agreement with observations in the frying industry. 
A key result from our model is that the lift-off time is largely controlled by the inflation of the vapour blanket. In fact, since the bubble inflation is so rapid, one can approximate the lift-off time as the time needed for first evaporation. Hence, as a proxy for the lift-off time, one can simply solve the initial heat diffusion problem (first regime) and find the time at which the temperature in the snack becomes uniformly equal to the evaporation temperature. If we do so, one of the key parameters that determines the lift-off time is the Nusselt number $\text{N}$, which is a measure of the heat conduction at the boundaries. In the literature the Nusselt number for snacks varies between $0.3$ and $1.3$. Therefore, in Figure \ref{nusselt} we plot the variation of approximate lift-off time with Nusselt number, where we also indicate some lift-off times calculated by solving the full numerical problem for $t^*$ such that $\rho(t^*)<1$.  The lift-off time is a monotonic decreasing function of $\text{N}$, as expected. In dimensional terms lift-off occurs for times between $0.5$ and $2.6$ seconds.

As a further motivation for our vapour blanket model, suppose instead we were to ignore the vapour blanket, and just solve the classic Stefan problem with Newton heating boundary conditions (i.e. $h=0$). In this case, we skip the second regime since there is no bubble inflation, and simply move from a heat diffusion regime to a quasi-steady regime. From Figure \ref{soln}e) we see that the density decay in the quasi-steady regime is much slower than that caused by bubble inflation. This results in a lift-off time closer to $t=1$, which in dimensional terms corresponds to more than $10$ seconds, and this is a factor of ten larger than experimental observations. Hence, this serves as a good indication that our vapour blanket model is accurate, and provides the essential ingredients to predict the lift-off time during frying.

Motivated by the above simulations, we now consider a further limiting case of the mathematical model called the quasi-steady limit. In this limit, we can further simplify the governing equations and find some analytical results that provide useful insight to the problem.

\section{Quasi-steady limit}\label{sec:quasi}

\begin{figure}
	\centering
	\begin{tikzpicture}[scale=0.5]
	\node at (-2,0) {\includegraphics[width=0.45\textwidth]{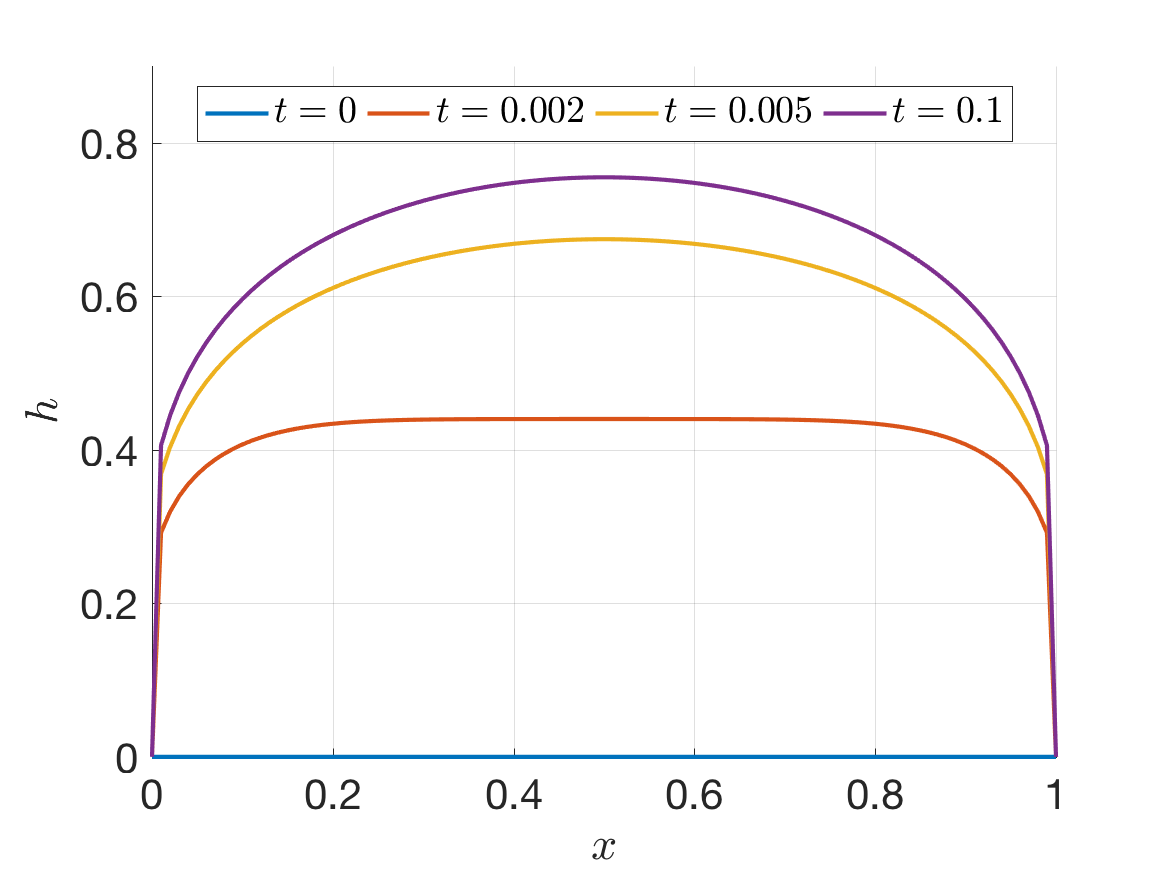}};
	\node at (-9.5,6.2) {a)}; 
	\node at (14,0) {\includegraphics[width=0.45\textwidth]{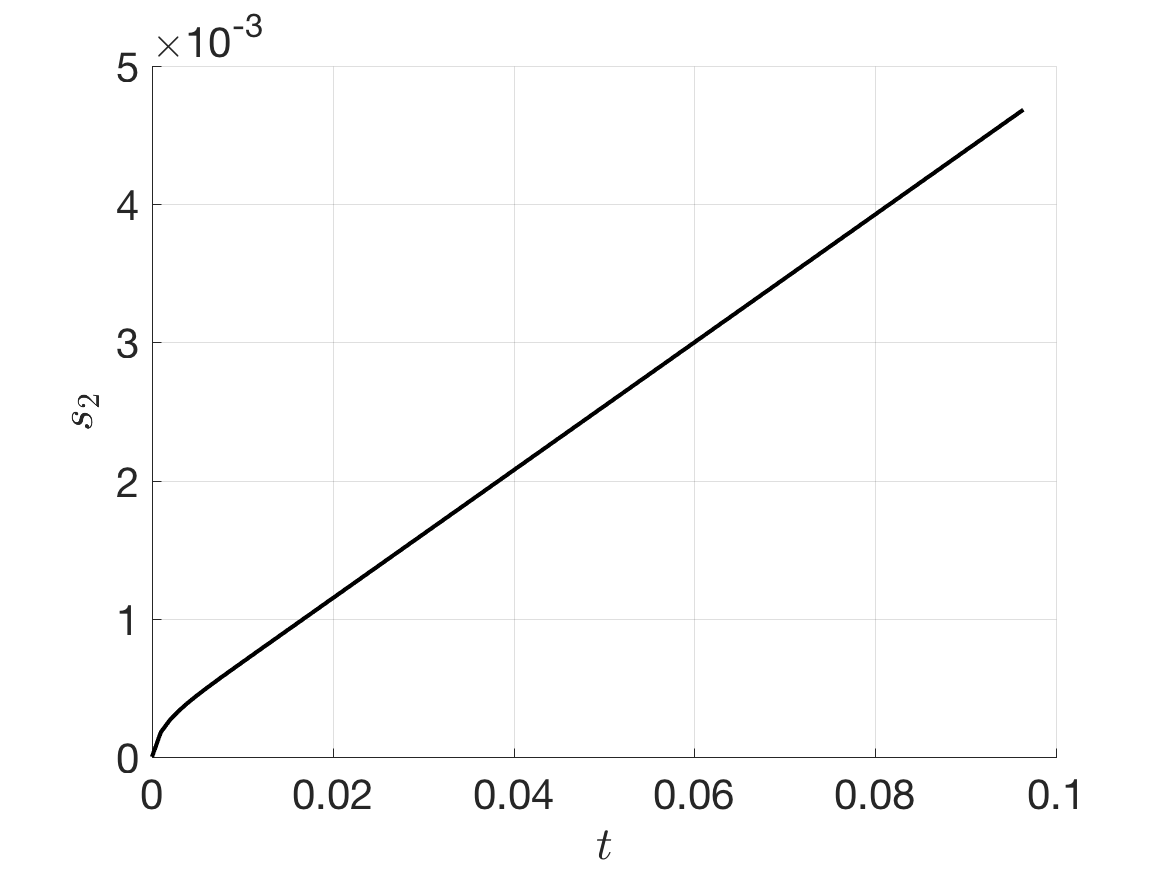}};
	\node at (7,6.2) {b)}; 
	\end{tikzpicture}
	  \caption{Solution to the quasi-steady approximation. (a) Evolution of the thin film ${h}$ at various times. (b) Evolution of the lower evaporation front ${s}_2({t})$.  }\label{quasifig1}
	
\end{figure}

The quasi-steady limit corresponds to when the thermal problem (\ref{region1eq})-(\ref{region3eq}) becomes independent of time except through the motion of the evaporation fronts. This limit, which is typical in such phase change problems, is a result of the fact that the Stefan number is large \cite{benham2016solidification, louro1986remarks, barry2008exact}. To study this limit, we restrict our attention to the second and third regimes of the above simulations. That is to say, we replace the above initial conditions of room temperature with initial conditions at the evaporation temperature $T(t=0)=0$. As before, we restrict our attention to the case where the evaporation fronts move uniformly, such that $s_1$, $s_2$ and $T_1$-$T_3$ are independent of $x$ and $y$.
Hence, the temperature in each region is given by
\begin{align}
    T_{i}&=A_{i}(t)z+B_{i}(t),
\end{align}
for some functions $A_{i}, B_{i}$, for $i=1-4$.
Applying the boundary conditions (\ref{nondimbc0})-(\ref{nondimbc5}), we obtain
\begin{align}
    T_{1}&=\frac{z-s_{1}}{1+1/\text{N}-s_{1}},\\
    T_{2}&=0,\\
    T_{3}&=\frac{s_{2}-z}{1/\text{N}+\bar{h}/\text{K}_2+s_{2}},\\
    %T_{4}&=\frac{s_{2}-z/\text{K}_2}{1/\text{N}+\bar{h}/\text{K}_2+s_{2}},\\
    \dot{s}_{1}&=\frac{-1}{1+1/\text{N}-s_{1}},\label{s1doteq}
    \\
    \dot{s}_{2}&=\frac{1}{1/\text{N}+\bar{h}/\text{K}_2+s_{2}}.\label{s2doteq}
\end{align}
The last equation (\ref{s2doteq}) contains the spatial average of $h$, which is found by solving the thin-film equation
\begin{equation}
    \frac{1}{\tau}\pdv{h}{t}=\dfrac{\beta}{3}\frac{\partial}{\partial {x}}\left( h^3\frac{\partial h}{\partial {x}}  \right)+\frac{1}{1/\text{N}+\bar{h}/\text{K}_2+s_{2}},\label{asymp1}
\end{equation}
together with the boundary conditions $h=0$ at $x=0,1$ from (\ref{flowbcs}). We can solve (\ref{s1doteq}) immediately, finding 
\beq
s_1=\frac{1}{\text{N}}\lb 1 + \text{N} - \sqrt{1 + 2\text{N}^2 {t}}\rb.\label{s1sol}
\eeq
The form of (\ref{s1sol}) reveals the classic $t^{1/2}$ similarity behaviour that is discussed for classic Stefan problems in the literature \cite{worster2000solidification, howison1988similarity}. The remaining unknowns ${h}$ and ${s}_2$ are found by solving the coupled system (\ref{s2doteq})-(\ref{asymp1}). In Figure \ref{quasifig1} we display the numerical solution to this system, calculated using the method of lines, as before. We see a fast early-time growth of the evaporation front ${s}_2$, causing a rapid inflation of the bubble over a timescale of around ${t}=0.01$. After this inflation period, the growth rate of ${s}_2$ is almost constant, and consequently the bubble shape reaches a steady state, which is consistent with Figure \ref{soln}.

\begin{comment}
Some analytic progress can be made by taking $\frac{1}{\tau}, \text{K}_2\rightarrow 0$. In this case the equation for $h$ simplifies to a quasistatic limit 
\begin{equation}
    \dfrac{1}{3}\frac{\partial}{\partial {x}}\left( h^3\frac{\partial h}{\partial {x}} \right)=-\dfrac{\text{K}_2}{h}.
\end{equation}
While this equation cannot be integrated, we can give an asymptotic solution valid near the
edge of the bubble
$$h\sim \left(\dfrac{25\text{K}_2}{6}\right)^{1/5}x^{2/5} \qquad x\rightarrow 0$$
where $x$ is a coordinate equal to zero on the boundary of the bubble and increasing perpendicular to the edge. 
\end{comment}

To understand the apparent steady state, let us consider the evolution equation for the lower evaporation front (\ref{s2doteq}). It is not immediately obvious that (\ref{s2doteq}) yields a constant growth rate solution. However, the non-dimensional conductivity ratio is very small $\text{K}_2\approx0.04$, and $s_2$ in Figure \ref{quasifig1} is also very small, suggesting that perhaps the variables $s_2$ and $h$ ought to be rescaled by $\text{K}_2$ appropriately. 
Since, for the steady state solution, we expect $h$ to be independent of time but dependent on space, and we expect the evaporation front to move at a linear growth rate, we seek a rescaling of the form
\begin{align}
s_2&=\text{K}_2^c (a+b t)+\mathcal{O}(\text{K}_2^{2c}),\\
h&=\text{K}_2^d H(x) + \mathcal{O}(\text{K}_2^{2d}),\label{weak}
\end{align}
for some unknown coefficients $a,b,c,d>0$. By inserting the above into (\ref{s2doteq})-(\ref{asymp1}), we can see that a steady state is only possible (to leading order) if we choose
\begin{align}
d-1+c=0,\\
4d=c,
\end{align}
which has solution $c=4/5$ and $d=1/5$. Taking the limit of small $\text{K}_2$, the resulting system of equations is
\begin{align}
b&=\frac{1}{\beta \bar{H}},\\
\frac{1}{3}\lb H^3 H_x \rb_x+\frac{1}{\beta \bar{H}}&=0,\label{steady2}
\end{align}
where $\bar{H}=\int_0^1 H \, \mathrm{d}x$ is the average film thickness, which is a constant in the steady state. 
%Note that the non-dimensional quantity $({R-1})/{\tau\beta}\approx 4.4$. 
We can solve (\ref{steady2}) to give $H$ in terms of its average value
\beq
H=\lb\frac{6x(1-x)}{\bar{H}\beta} \rb^{1/4}. \label{steady3}
\eeq
and $\bar{H}$ is found by integrating (\ref{steady3}), which gives
\beq
\bar{H}=\lb \frac{\Gamma (5/4)}{\Gamma(3/4)}\lb\frac{8\pi^2}{27\beta}\rb^{1/4}  \rb^{4/5},
\eeq
where $\Gamma$ is the Euler Gamma function. Note that the above is only valid for times much smaller than $t\approx \text{K}_2^{-4/5}\approx13$. However, the snack frying process all takes place within $0\leq t\leq 1$, so this is acceptable. Note also that the linear behaviour of $s_2$ with respect to time is different from the square root behaviour of $s_1$ observed in (\ref{s1sol}). Hence, the vapour blanket completely changes evaporation at the lower boundary.

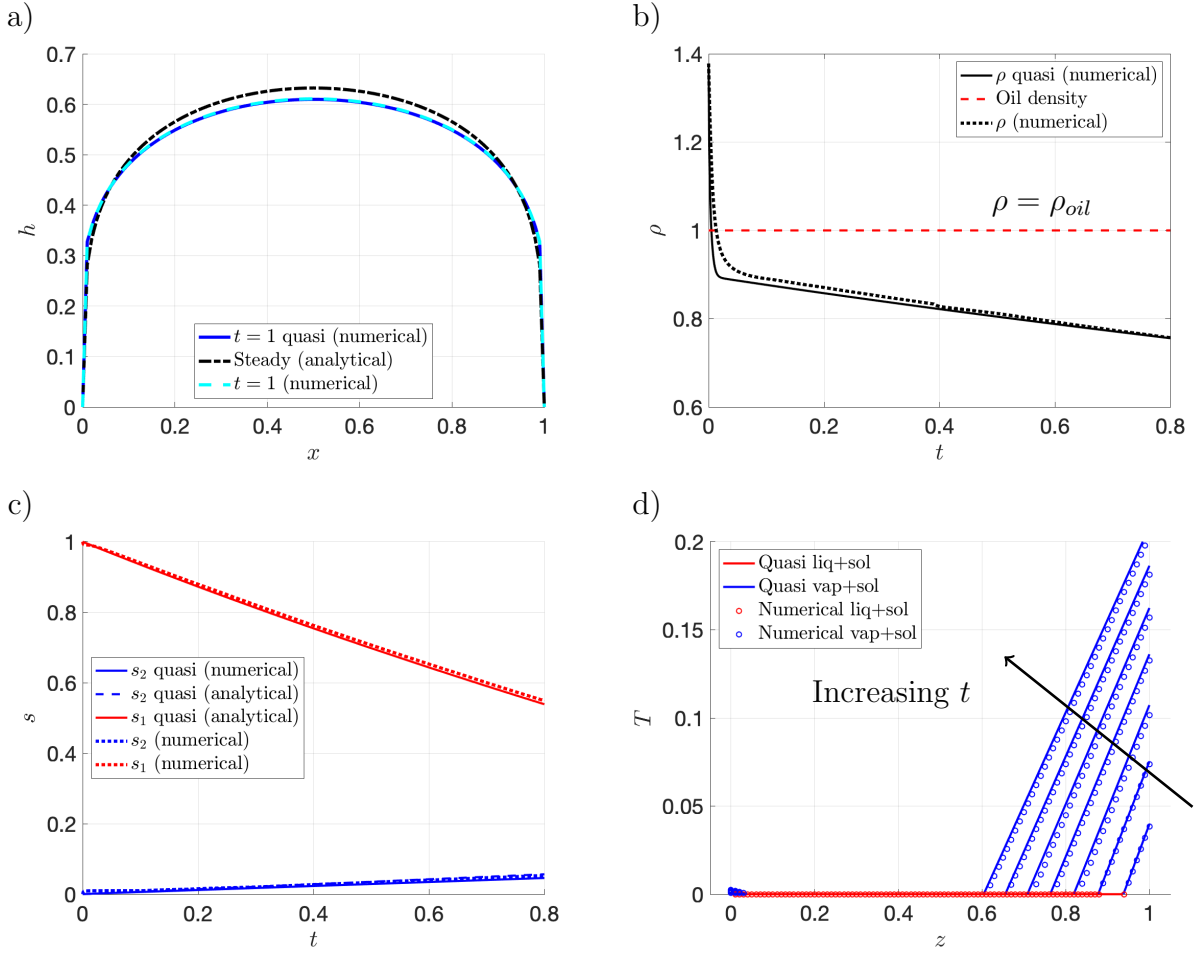
\begin{figure}
    \centering
    
       \begin{tikzpicture}[scale=0.5]
       \node at (0,0) {
    \includegraphics[width=0.45\textwidth]{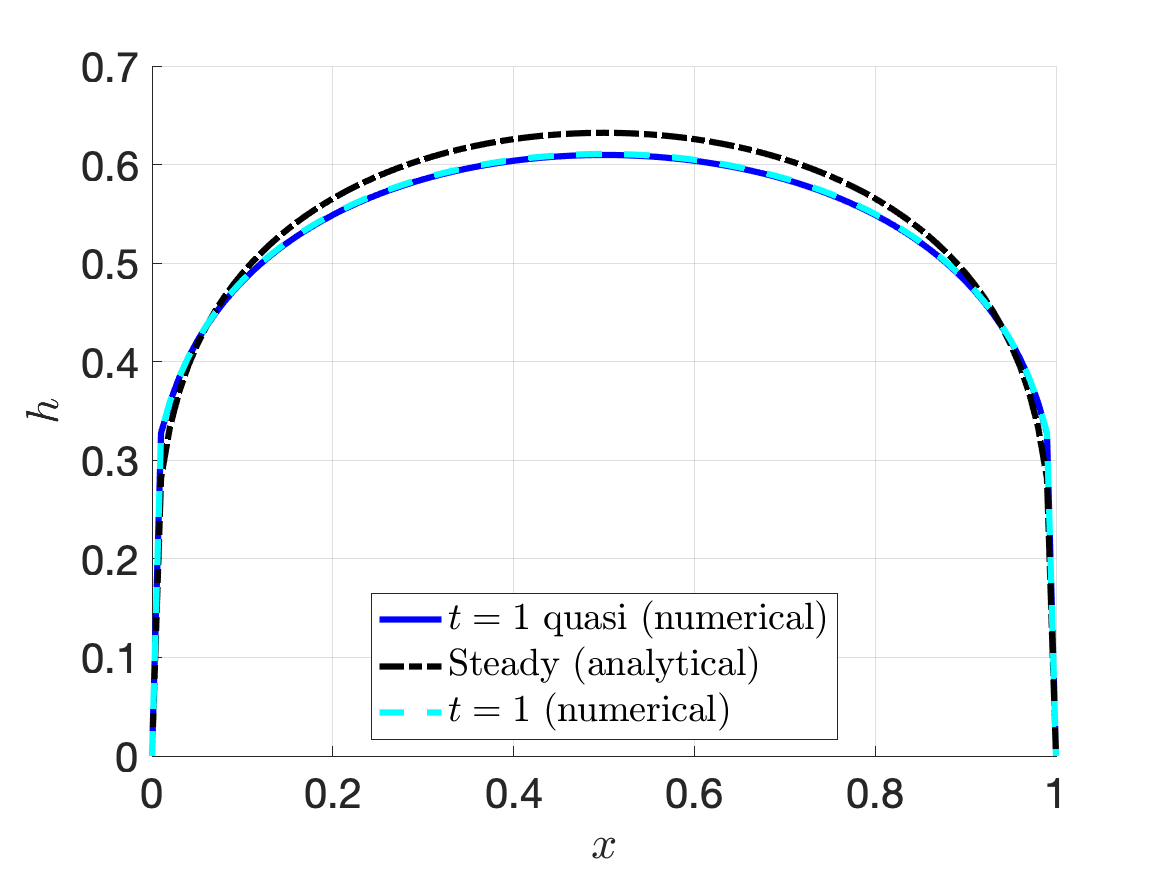}};
     \node at (-7.5,6) {a)}; 
     \end{tikzpicture}
    %\caption{Spatial profile of $h$ at $t=0.01$.}
    %\label{fig:q}
%\end{figure}
%\begin{figure}
    %\centering
    \begin{tikzpicture}[scale=0.5]
    \node at (0,0) {\includegraphics[width=0.45\textwidth]{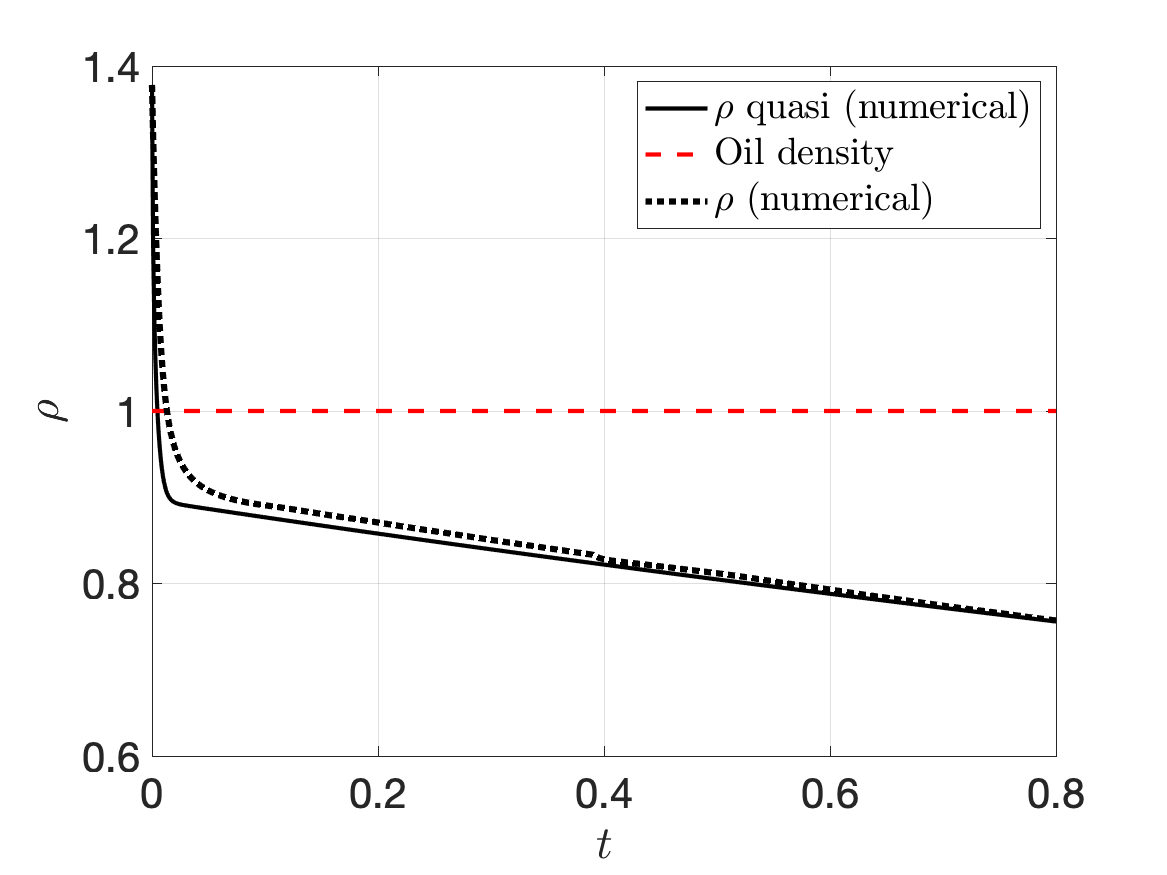}};
    \node at (3,1) {$\rho=\rho_{oil}$};
    \node at (-7.5,6) {b)}; 
    \end{tikzpicture}\\
    %\caption{Temporal profile of $h$ at $x=0.5$ until $t=0.1$.}
    %\label{fig:q1}
%\end{figure}
%\begin{figure}
    %\centering
      \begin{tikzpicture}[scale=0.5]
      \node at (0,0) {
    \includegraphics[width=0.45\textwidth]{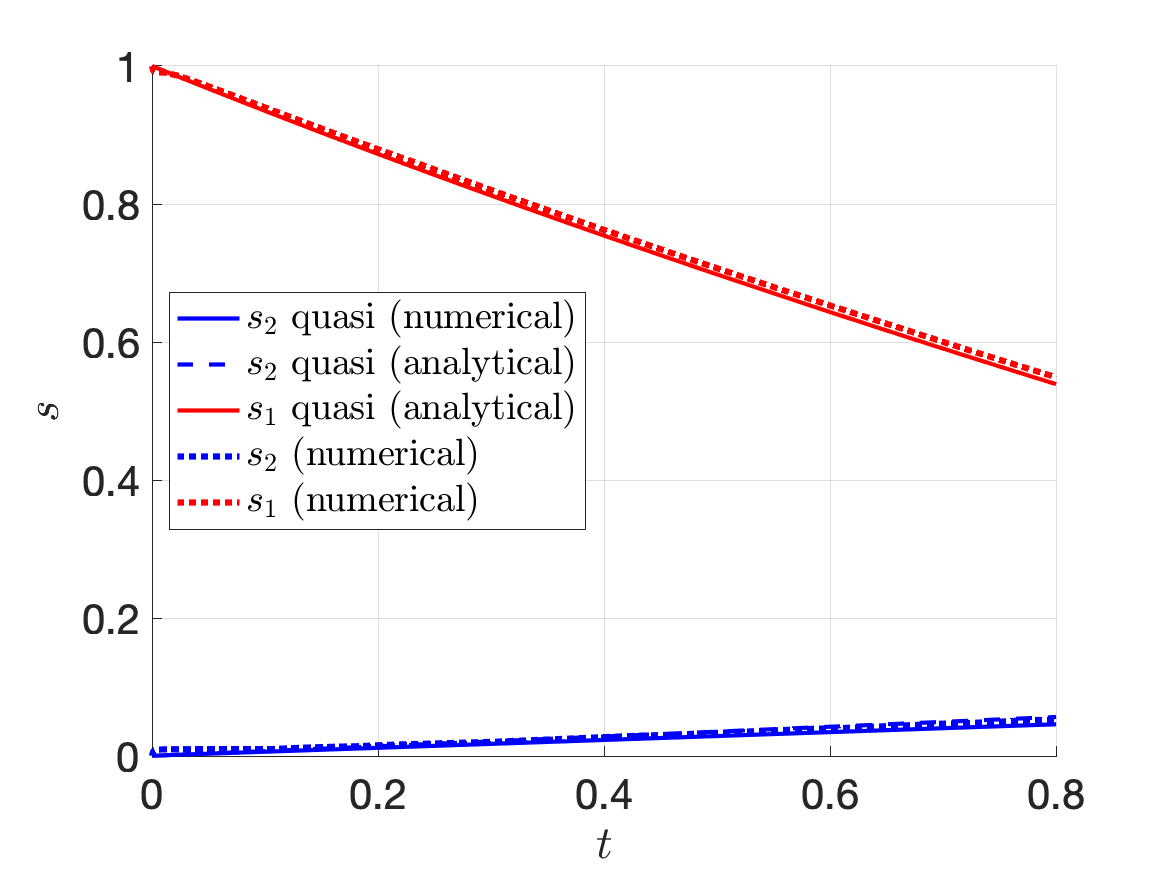} };
  \node at (-7.5,6) {c)}; 
  \end{tikzpicture}
    \begin{tikzpicture}[scale=0.5]
    \node at (0,0) {\includegraphics[width=0.45\textwidth]{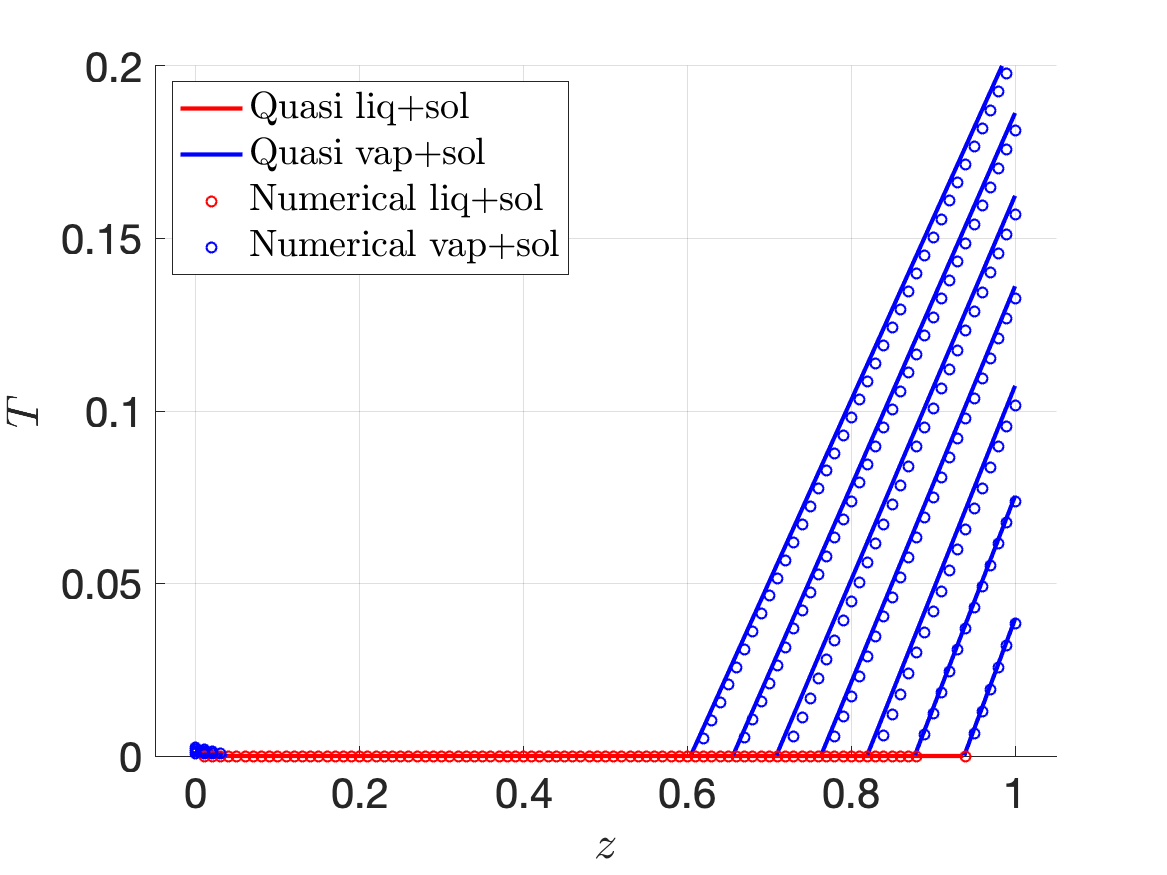}};
    \draw[line width =1,->] (7,-2)--(2,2);
    \node at (-1,1) {Increasing ${t}$};
     \node at (-7.5,6) {d)}; 
    \end{tikzpicture}
    \caption{Long time results from the quasi-steady approximation. (a) Evolution of the thin film ${h}$ at various times, compared to the analytical solution for the steady state. (b) Density ${\rho}({t})$ as a function of time, indicating the lift-off density $\rho=\rho_{oil}$. (c) Evolution of the lower Stefan boundary ${s}_1({t})$, compared to the analytical solution for the steady state $\dot{{s}}_1=\text{K}_2^{4/5}/\bar{H}$. (d) Temperature profiles ${T}({z},{t})$ at different times between ${t}=0$ and ${t}=1$, indicating liquid and vapour regions.   }\label{quasifig2}
   
\end{figure}

In Figure \ref{quasifig2} we display a comparison of the results from the quasi-steady limit, including the steady state, to the original numerical solution from Figure \ref{soln}. For the comparison, we look at the long-time evolution of the film thickness, the density, the evaporation fronts and the temperature within the snack. We see that in all cases there is close agreement between the numerical solution to the full problem, the quasi-steady solution and the steady state. There is a slight discrepancy ($\sim 5\%$) for the steady state solution to the thin film, and this can be explained by the asymptotic approximation (\ref{weak}). This discrepancy could be mitigated by going to higher order terms in the asymptotic expansion. 

There is also a slight disagreement ($\sim 5\%$) between the early-time density predictions of the numerical solution to the full problem and the quasi-steady solution. 
%In particular, the full numerical solution shows a non-monotonic behaviour for $\rho$, whereas the quasi-steady solution is monotonic decreasing with time. 
This can be explained by the way in which we calculate the speed of the lower evaporation front $\dot{s}_2$, which largely controls the density at early times via the inflation of the vapour blanket.  
In the quasi-steady approximation we calculate the evaporation front $s_2$ using a numerical discretisation scheme in time to solve (\ref{s2doteq}), with time step $\delta t=2\times 10^{-7}$, providing very smooth results. On the other hand, in the numerical solution to the full problem, since we calculate the temperature using the enthalpy method, which does not require tracking the position of the fronts, the evaporation front is calculated by finding the grid point that separates liquid and gas phases. Since the grid spacing is finite, this leads to non-smooth step changes in $s_2$ and spikes in the time-derivative of $s_2$, which we have attempted to smooth using a damping method. Nevertheless, even with a time step $\delta t=2\times 10^{-7}$, this produces inevitable error associated with the inflation of the vapour blanket, and this is reflected in the slight disagreement for the density prediction at early times. 

Closer agreement can be attained with an even smaller spatial discretisation, but due to the explicit discretisation method, this results in lengthy computation times. Hence, interestingly the quasi-steady solution, though it only applies to an asymptotic limit, is generally more accurate than the numerical solution to the full problem. Since the critical time of interest is the lift-off time, which still shows close agreement between these two approaches, we do not consider this discrepancy to be very important.

Finally, in Figure \ref{quasifig2} c,d) we display a comparison of the predictions of the evaporation fronts and the temperature. On a macroscopic level, there is very close agreement, and in particular the steady state solution performs remarkably well. After a time of $t=1$, or 10 seconds in dimensional terms, nearly half the liquid in the snack has evaporated and the density has dropped by a factor of around 2. The overall thickness of the vapour blanket is nearly equal to the total width of the snack, which is also consistent with experimental observations.

\section{Conclusions}\label{sec:conclusion}

We considered a mathematical model of potato-snack frying in order to obtain an estimate for the lift-off time of the snack from the conveyor belt. To that end, we modelled the frying process as a Stefan problem with two propagating evaporation fronts where the liquid in the dough turns into vapour and decreases the density of the snack. In addition, a key feature in our model is the presence of a vapour blanket that forms underneath the snack as liquid evaporates. The moving vapour fronts and the vapour blanket were assumed to be the two main mechanisms for density reduction of the snack and, therefore, its eventual lift-off from the belt. Numerical results of the full system, using the enthalpy method, revealed that, indeed, both of these mechanisms were essential to predict a physically realistic lift-off time of the order of a second. Furthermore, we considered a simplified quasi-steady model due to the large Stefan number. Numerical solutions to the reduced problem agreed very well with solutions to the full system and thus allow for a computationally cheaper way to investigate properties of our model and, in particular, the lift-off time.

One of the key dimensionless parameters that emerged as part of our analysis was the Nusselt number $\text{N}$, which is the ratio between heat transfer at the snack boundary and heat conduction in the snack interior. We investigated how changing $\text{N}$ affects the lift-off time of the snack. This is important to snack manufacturers since changing the dough, for example, can change the material properties and hence the parameters of the system. Having the dependence of the lift-off time on these parameters is useful in determining the optimal cooking strategies.

%% Begin Stacie Remove
%% For future work, the effect of the initial detachment of the dough from the substrate could be included in our model. This would require a careful consideration of surface-tension and elasticity effects of the dough. What is more, the oil itself flows with a given speed, which may also contribute to the detachment of the crisp, especially when the conveyor belt is slightly inclined. The effect of the belt mesh on the detachment process could also be investigated, including a comparison of different designs.
%% End Stacie Remove
%% Begin Stacie Add
To further improve the prediction of lift-off time for the snack there should be a consideration of other forces. These could include interfacial tension between the snack and belt as well as the peeling energy required to overcome the dough elasticity. As a result the orientation of the snack on the belt and indeed the belt design and material could have a further impact on the lift-off time of the snack.
%% End Stacie Add

\section*{Acknowledgments}
The authors would like to aknowledge the 138th European Study Group with Industry which was held in Bath, 16-20 July 2018, and jointly hosted by the University of Bath and University of Bristol. The authors are also thankful to all the participants of the ESGI project for useful discussions: Peter Baddoo (University of Cambridge, UK), Sean Bohun (University of Ontario, Canada), Stephen Cowley (University of Oxford, UK), Helen Fletcher (University of Oxford, UK), Harry Reynolds (University of Oxford, UK), and Chris Sear (University of Cambdrige, UK). 
Special thanks to PepsiCo for proposing the problem and funding the work throughout the ESGI, and for their collaboration during and after the study group was held. The views expressed in this article are those of the authors and do not necessarily reflect the position or policy of PepsiCo, Inc. Kris Kiradjiev, Thomas Babb and Raquel Gonzalez-Farina would also like to acknowledge the support of EPSRC Center for Doctoral Training in Industrially Focused Mathematical Modelling (EP/L015803/1).

%% The Appendices part is started with the command \appendix;
%% appendix sections are then done as normal sections

%\pagebreak
%\begin{comment}

%\end{comment}

%% \label{}

%% References
%%
%% Following citation commands can be used in the body text:
%% Usage of \cite is as follows:
%%   \cite{key}          ==>>  [#]
%%   \cite[chap. 2]{key} ==>>  [#, chap. 2]
%%   \citet{key}         ==>>  Author [#]

%% References with bibTeX database:

%\bibliographystyle{model1-num-names}
\bibliographystyle{plain}
\bibliography{references.bib}

%% Authors are advised to submit their bibtex database files. They are
%% requested to list a bibtex style file in the manuscript if they do
%% not want to use model1-num-names.bst.

%% References without bibTeX database:

% \begin{thebibliography}{00}

%% \bibitem must have the following form:
%%   \bibitem{key}...
%%

% \bibitem{}

% \end{thebibliography}

\end{document}